\documentclass[aps,prr,twocolumn,nofootinbib,superscriptaddress,showkeys]{revtex4-1}

\usepackage{color}
\usepackage{graphicx}
\usepackage{dcolumn}
\usepackage{bm}
\usepackage{lipsum}
\usepackage{chngcntr}
\usepackage[mathlines]{lineno}
\usepackage[english]{babel}
\usepackage[latin1]{inputenc}
\usepackage[bottom=2cm,top=2cm, left=1.5cm, right=1.5cm]{geometry}
\usepackage{booktabs}
\usepackage[unicode=true, bookmarks=true, bookmarksnumbered=false, bookmarksopen=false, breaklinks=false, pdfborder={0 0 1}, backref=false, colorlinks=false]{hyperref}

\usepackage{natbib}
\usepackage{color,amsmath}

\hypersetup{pdftitle={Magnetoelectric domain engineering from micrometer to {\AA}ngstr{\o}m scales}, pdfauthor={Marcela Giraldo, Arkadiy Simonov, Hasung Sim, Ahmed Samir Lotfy, Martin Lilienblum, Lea Forster, Elzbieta Gradauskaite, Morgan Trassin, Je-Geun Park, Thomas Lottermoser, Manfred Fiebig}, pdfkeywords={multiferroics, SHG imaging}}

\begin{document}

\title{Magnetoelectric domain engineering from micrometer to {\AA}ngstr{\o}m scales}

\author{Marcela Giraldo}
\affiliation{Department of Materials, ETH Zurich, 8093 Zurich, Switzerland.}

\author{Arkadiy Simonov}
\affiliation{Department of Materials, ETH Zurich, 8093 Zurich, Switzerland.}

\author{Hasung Sim}
\affiliation{Department of Physics and Astronomy, Seoul National University, South Korea.}

\author{Ahmed Samir Lotfy}
\affiliation{Quantum Solid State Physics, KU Leuven, 3001 Leuven, Belgium}

\author{Martin Lilienblum}
\affiliation{Department of Materials, ETH Zurich, 8093 Zurich, Switzerland.}

\author{Lea Forster}
\affiliation{Department of Materials, ETH Zurich, 8093 Zurich, Switzerland.}

\author{Elzbieta Gradauskaite}
\affiliation{Laboratoire Albert Fert, CNRS/Thales.}

\author{Morgan Trassin}
\affiliation{Department of Materials, ETH Zurich, 8093 Zurich, Switzerland.}

\author{Je-Geun Park}
\affiliation{Department of Physics and Astronomy, Seoul National University, South Korea.}

\author{Thomas Lottermoser}
\affiliation{Department of Materials, ETH Zurich, 8093 Zurich, Switzerland.}

\author{Manfred Fiebig}
\email{manfred.fiebig@mat.ethz.ch}
\affiliation{Department of Materials, ETH Zurich, 8093 Zurich, Switzerland.}

\date{\today}

\begin{abstract}

The functionality of magnetoelectric multiferroics depends on the formation, size, and coupling of their magnetic and electric domains. Knowing the parameters guiding these criteria is a key effort in the emerging field of magnetoelectric domain engineering. Here we show, using a combination of piezoresponse-force microscopy, non-linear optics, and x-ray scattering, that the correlation length setting the size of the ferroelectric domains in the multiferroic hexagonal manganites can be engineered from the micron range down to a few unit cells under the substitution of Mn$^{3+}$ ions with Al$^{3+}$ ions. The magnetoelectric coupling mechanism between the antiferromagnetic Mn$^{3+}$ order and the distortive-ferroelectric order remains intact even at substantial replacement of Mn$^{3+}$ by Al$^{3+}$. Hence, chemical substitution proves to be an effective tool for domain-size engineering in one of the most studied classes of multiferroics.

\end{abstract}

\maketitle

\begingroup\renewcommand\thefootnote{\textsection}

\endgroup

\section{INTRODUCTION}
Multiferroics are commonly understood as materials with a coexistence of magnetic and ferroelectric order. Important criteria determining their practical value are the magnetoelectric coupling strength between the coexisting orders, the ordering temperatures defining the range of operation, and also the formation, size, and coupling of the magnetic and electric domains \cite{Matsubara2015, Leo2018, Giraldo2021, Fiebig2002}. In particular, the size of the domains sets inherent spatial limitations for magnetoelectric applications. In multiferroic materials-based memory elements, for example, this size would define the expansion of the magnetoelectric data unit and, hence, the processing or storage capacity \cite{Bibes2008}. 

Because of the central role of the domains for the functionality of a multiferroic material, developing pathways for domain engineering is vital for effectively implementing the magnetoelectric multiferroics in device-like environments. There are many well-established procedures for steering the domain size through poling procedures, annealing, strain, and other external criteria \cite{Volk2008, Gradauskaite2023, Strkalj2019, Manz2016}. A hitherto little explored parameter is chemical substitution. By replacing one of the ions contributing to the formation of an ordered state, the balance of long-range interactions stabilizing this state is perturbed with consequences for domain formation \cite{Huang2014}.

An indication that chemical substitution can be a very valuable degree of freedom here was provided by experiments on BiFeO$_3$, the only robust room-temperature multiferroic to date. It turned out that replacement of Bi$^{3+}$ by La$^{3+}$ lowers the coercive field and creates a richer, more versatile ferroelectric domain pattern than the non-substituted compound \cite{Muller2021, Manipatruni2019, Huang2020}. Expanding this approach to other materials with coexisting magnetic and electric order permits to further explore the parameter space of chemical substitution and its underlying mechanisms in the domain formation in multiferroics.

A more extended family of materials that shows promising signs toward domain engineering are the multiferroic hexagonal manganites, $R$MnO$_3$ with $R={\rm Sc, Y, In, Dy-Lu}$. The compounds exhibit a rich variety of magnetoelectric and transport effects at the level of their domains and domain walls. Specifically, the ferroelectric order exhibits topologically protected domains \cite{Choi2010, Jungk2010, Lilienblum2015}, domain walls with anisotropic conductance \cite{Meier2012}, and antiferromagnetic domains exhibiting different types of domain walls that are either coupled to the ferroelectric domain walls or not \cite{Giraldo2021, Fiebig2002, Artyukhin2014}. 

The size of both the ferroelectric and the antiferromagnetic domains can be controlled by appropriate thermal annealing procedures \cite{Griffin2012, Meier2017, Fiebig1998}. Thermal quenching procedures are inconvenient for domain-size engineering, however, and therefore alternative mechanisms like chemical substitution may be explored \cite{Polat2018, Fu2016, Hassanpour2016, Sekhar2005, Holstad2018, Park2021, Huang2014, Nugroho2007, Choi2017, Cho2020, Sim2018, Zhou2005, Park2009}. In ErMnO$_3$, replacing Er$^{3+}$ with Ca$^{2+}$ or Zr$^{4+}$ reduces or enhances the bulk conductivity, respectively. Replacing In$^{3+}$ by Ga$^{3+}$ in InMnO$_3$ preserves the archetypal distortive-dielectric domain configuration, but yields two types of order, where one is ferroelectric, and the other one is not \cite{Huang2014}. Substitution of Al$^{3+}$ at the Mn$^{3+}$ site in YMnO$_3$ showed that the replacement reduces the lattice distortion promoting the improper ferroelectric order \cite{Sim2018}. It would now be very interesting to see to what extent chemical substitution in the hexagonal $R$MnO$_3$ system influences the size of the ferroelectric or antiferromagnetic domains, and also the coupling between the different types of domains, but such an investigation is still pending.

Here we show for hexagonal YMnO$_3$, the most studied representative of the $R$MnO$_3$ series, that substitution of Mn$^{3+}$ by Al$^{3+}$ is a handle to tune the correlation length determining the size of the ferroelectric domains, with an expansion ranging from a few $\mu$m to a few unit cells and, hence, across three orders of magnitude. This surpasses the range accessible by thermal annealing procedures. Comparison of Al$^{3+}$- and Ga$^{3+}$-substituted YMnO$_3$ reveals that anisotropic lattice strain is responsible for the observed variation of domain size. We also find that even at high degrees of substitution, the coupling mechanism between the ferroelectric and antiferromagnetic domains remains intact, thus preserving the magnetoelectric coupling. Methodically, we investigate the size of the magnetic and electric domains in a combination of piezoresponse-force microscopy (PFM), optical second harmonic generation (SHG), and single-crystal x-ray diffraction.

\section{MATERIALS AND METHODS} \label{methods}

\textbf{The multiferroic hexagonal manganites:} The hexagonal $R$MnO$_3$ compounds represent a family of multiferroics where the ferroelectric and the antiferromagnetic orders emerge independently. The unit cell is formed by planes of $R^{3+}$ ions alternating with sheets of corner-sharing MnO$_{5}$ octahedra along the hexagonal $c$-axis. The ferroelectric order is distortively driven and thus improper. At a temperature between 1250 and 1700~K \cite{Lilienblum2015, LilienblumPhD2015} a lattice trimerization occurs that induces a spontaneous polarization along the hexagonal $c$-axis as secondary effect. The transition lowers the symmetry from the paraelectric P6$_{3}$/mmc to the ferroelectric P6$_{3}$cm space group. The lattice distortion and trimerization lead to a network of topologically protected ferroelectric domains, where sets of six domains of alternating polarization converge at a common point \cite{Choi2010, Jungk2010, Lilienblum2015}. Antiferromagnetic order of the Mn$^{3+}$ ions emerges at temperatures between 65 and 120~K. It consists of a triangular arrangement of Mn$^{3+}$ spins in the $ab$-planes, which are weakly coupled along the $c$-axis. Below 10~K, magnetic rare-earth order may occur, but here we focus on YMnO$_3$ in which rare-earth ordering is absent. This compound also stands out as the most studied representative of the hexagonal $R$MnO$_3$ family.

\textbf{Sample Preparation:} Single crystals of YMn$_{1-x}$Al$_{x}$O$_3$ with $x={\rm 0, 0.05, 0.10, 0.15, 0.20, 0.25}$ and YMn$_{1-x}$Ga$_{x}$O$_3$ with $x={\rm 0, 0.25, 0.50}$ were grown by the floating-zone technique using growth parameters as described elsewhere \cite{Sim2018, Park2009}. The crystals were oriented using an Oxford Xcalibur PX Ultra single-crystal x-ray diffractometer. Slices were cut perpendicular to the $c$-axis using a diamond saw and thinned to $70-80$~$\mu$m by lapping. For YMn$_{1-x}$Al$_{x}$O$_3$ with $x<0.15$ this was done using a slurry of Al$_2$O$_3$ powder with a grain size of 9~$\mu$m. YMn$_{1-x}$Ga$_{x}$O$_3$ samples were lapped with Logitech SiC powder, 1200 grit. All samples were chemo-mechanically polished with Eminess Ultra-Sol\texttrademark R2EX solution to reveal surfaces with a root-mean-square roughness of $\leq 3$~nm. YMn$_{1-x}$Ga$_{x}$O$_3$ samples with $x=0.20,0.25$ were lapped and polished using Al$_2$O$_3$ Thorlabs sheets with subsequent grain sizes of 5~$\mu$m, 3~$\mu$m, 1~$\mu$m, and 0.3~$\mu$m.

\begin{figure*}[ht!] 
\centering
\includegraphics[scale = 1 ]{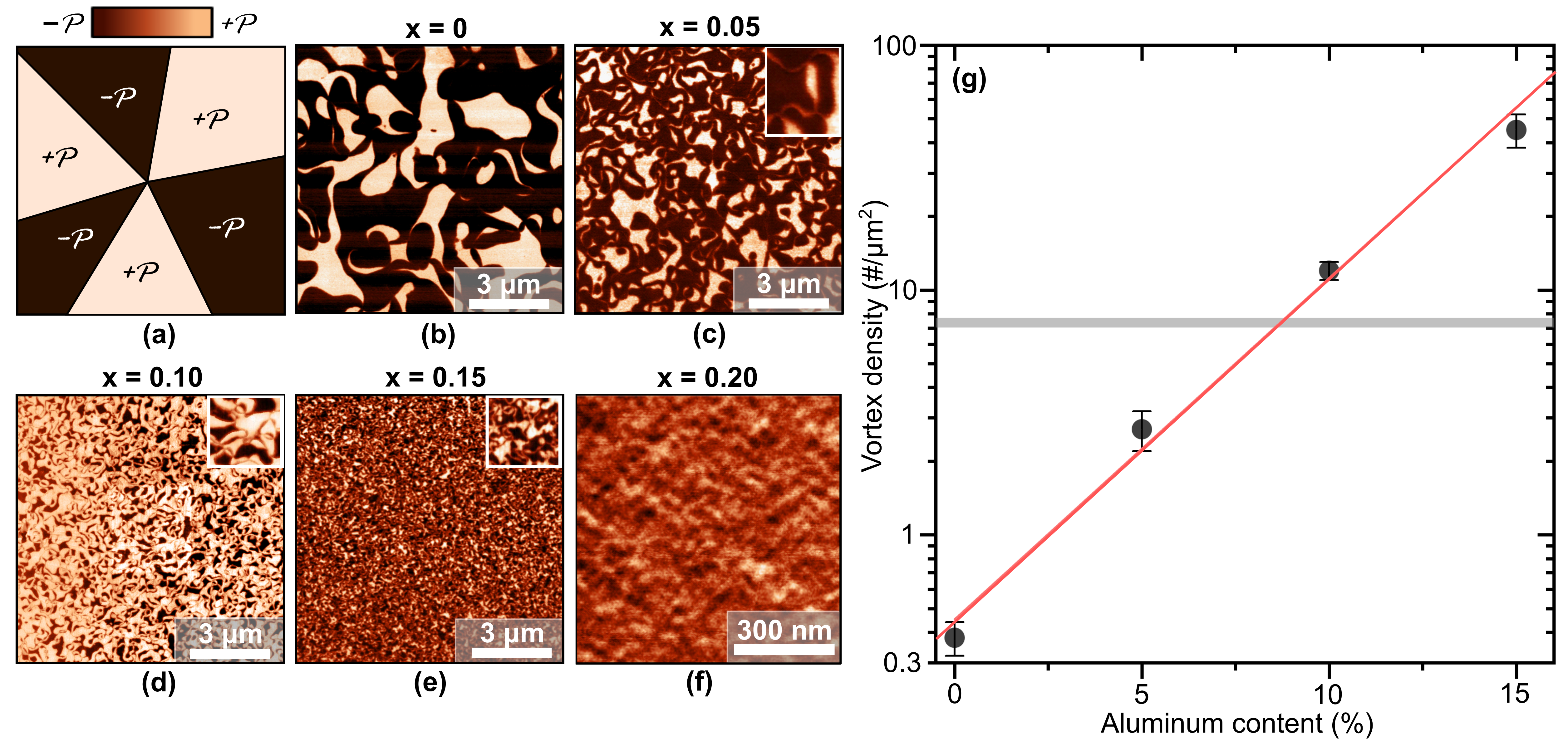}
\caption{\label{fel-al} Distribution and density of ferroelectric domains in the YMn$_{1-x}$Al$_{x}$O$_{3}$ series of the multiferroic hexagonal manganites. (a) Schematic of the six-fold vortex-like arrangement of ferroelectric domains with alternating $\pm\mathcal{P}$ polarization states. (b -- f) Room-temperature out-of-plane PFM scans on a $c$-oriented area of $10\times 10$~$\mu$m$^{2}$ for different substitution $x$. Insets show magnified regions of $1\times 1$~$\mu$m$^{2}$. (g) Density of six-fold vortex-like domain meeting points on the samples from (b -- e). The red line represents a linear fit. The gray horizontal bar indicates the largest vortex density achieved in YMnO$_{3}$ by thermal quenching \cite{Griffin2012, Meier2017}. It is exceeded by an order of magnitude with chemical substitution.}
\end{figure*}

\textbf{PFM:} We used PFM for the investigation of the as-grown domain structure in $c$-oriented YMn$_{1-x}$Al$_{x}$O$_3$ and YMn$_{1-x}$Ga$_{x}$O$_3$ samples. The out-of-plane amplitude of the PFM signal was recorded with a conductive tip in contact mode. Measurements were conducted with an NT-MDT NTEGRA Prima scanning probe microscope with Pt-coated HQ:NSC35/Pt (MikroMasch) tips. Samples were mounted on a metallic plate using silver paint as a contact electrode. Measurements were carried out at room temperature using Stanford Research SR 830 lock-in amplifiers and applying 5~V peak-to-peak AC modulation at 70~kHz to the rear of the samples. The ferroelectric domain patterns in the $c$-oriented YMn$_{1-x}$Al$_{x}$O$_3$ samples were scanned using resonant PFM to achieve a higher signal-to-noise ratio and therefore a higher sensitivity \cite{Gruverman2019, Kalinin2007, Kalinin2007_2, LilienblumPhD2015}. For electric poling toward opposite domain states in the YMn$_{0.80}$Al$_{0.20}$O$_3$ samples, voltages of $\pm$ 250~V were applied between the tip and the rear electrode. A Keithley 6517B electrometer was used as the voltage source.

\textbf{SHG:} SHG denotes the frequency doubling of light in a material and is described by the relation $P_i(2\omega)=\epsilon_0\chi_{ijk}E_j(\omega)E_k(\omega)$. Here, $\vec{E}(\omega)$ is the electric field of the incident fundamental light, and $\vec{P}(2\omega)$ is the frequency-doubled polarization, where the latter is the source of the SHG light emitted from the crystal. Further, $\epsilon_0$ is the vacuum permittivity, and $\hat{\chi}$ is the nonlinear susceptibility tensor. Because of its sensitivity to symmetry, SHG is an ideal tool to detect and distinguish the inherently symmetry-breaking electric and magnetic orders in multiferroics. The spectrally and spatially resolved investigation of the ferroelectric and antiferromagnetic orders of the hexagonal manganites is well understood \cite{Giraldo2021, Sa2000, Pisarev2000, Hanamura2003, Fiebig2005b}. In short, there are SHG contributions coupling to the ferroelectric order parameter $\mathcal{P}$ and to the product of ferroelectric and antiferromagnetic order parameters, $\mathcal{PL}$. In the electric-dipole approximation of the light field, SHG is permitted only in non-centrosymmetric systems, and both $\mathcal{P}$ and $\mathcal{PL}$ couple to the loss of inversion symmetry by the ferroelectric order. Earlier experiments revealed that the most pronounced SHG signals from $\mathcal{P}$ are obtained with the SHG susceptibility component $\chi_{zzz}$ at SHG photon-energies around 1.7 or 2.5~eV  \cite{Lottermoser2002}. For addressing $\chi_{zzz}$, the $c$-oriented samples has to be tilted (here by about $30^{\circ}$) against the incoming laser beam. The most pronounced SHG signals from $\mathcal{PL}$ are obtained with the SHG susceptibility component $\chi_{yyy}$ at 2.45~eV as SHG photon energy. With the $c$-oriented samples, this component is best accessed in normal incidence. 

Our $c$-oriented, chemically substituted YMnO$_3$ samples were probed at room temperature or 20~K in an optical transmission setup described elsewhere \cite{Fiebig2005b}. As the light source, we used a Coherent Legend Elite Duo laser system coupled to an optical parametric amplifier (Coherent OPerA Solo). The pulse length was 120~fs, and pulses were emitted at 1~kHz repetition rate. Samples were probed with pulses of 15~$\mu$J at a SHG photon energy of 2.45~eV. The frequency-doubled light was collected using a microscope objective (magnification 3.5, numerical aperture 0.1). As the detector, we used a Horiba Jobin Yvon back-illuminated deep-depletion digital camera with a near-infrared detector chip of $1024 \times 256$ pixels. Thermal noise in the detector was reduced by cooling the camera with liquid nitrogen. We used a Janis ST-500 optical micro-cryostat for the imaging measurements below the antiferromagnetic ordering temperature. Temperature was stabilized with a precision of $\pm 0.05$~K with the use of a Lakeshore 331 controller.

\textbf{Synchrotron diffraction:}  Single-crystal diffuse-scattering experiments were performed at the beamline ID28 at the European Synchrotron Radiation Facility (ESRF). The datasets were measured at a wavelength $\lambda = 0.7839$~\AA{} using the Pilatus3 1M detector with a silicon layer thickness of 450~$\mu$m. YMn$_{1-x}$Al$_{x}$O$_3$ single crystals of appropriate size were cleaved with a scalpel. No special surface treatment was performed, as we did not observe any significant surface-scattering background. Crystal orientation was determined using the program XDS \cite{XDS}, and reciprocal maps were reconstructed using the program Meerkat. 

As mentioned, ferroelectricity in YMn$_{1-x}$Al$_{x}$O$_3$ is driven by trimerization of the unit cell and is thus improper. The paraelectric P6$_3$/mmc phase has the lattice parameters $a=b=3.58$~{\AA}, $c=11.3$~{\AA}, whereas the ferroelectric P6$_3$cm phase features a three times larger unit cell with $a=b=6.19$~{\AA}, $c=11.3$~{\AA}. From the structural point of view, the trimerization is driven by the tilts of the MnO$_5$ bipyramids and the displacement of Y atoms along the $c$ axis. This means that the amplitude of the ferroelectric displacement is proportional to the intensity of reflections satisfying the condition $h-k \neq 3n$, where the correlation length can be estimated from the width of those reflections. We modelled diffuse scattering using a custom script, convolving the reflections associated with the trimerized-ferroelectric structure with a Gaussian function. Three parameters were refined for each model: an overall scaling factor, the Gaussian width along the $c$ direction, and the Gaussian width along the symmetry-equivalent $a$ and $b$ directions.


\section{RESULTS AND DISCUSSION}

\subsection{Ferroelectric order and domain size}

Figure~\ref{fel-al} shows the ferroelectric domain pattern at room temperature in the $ab$-plane of four YMn$_{1-x}$Al$_{x}$O$_3$ samples with increasing Al$^{3+}$ substitution. This is complemented by an analysis of the associated density of the six-fold vortex-like ferroelectric domain-meeting points. The most striking impression of Fig.~\ref{fel-al} is the pronounced dependence of the domain size on the Al$^{3+}$ content $x$. In all images where domains are resolved, we observe the characteristic vortex-like distribution of ferroelectric domains described in Section~\ref{methods}. However, the lateral extension of the domains changes from about 2~$\mu$m at $x=0$ to 150~nm at $x=0.15$, and at $x=0.20$, domains are no longer resolved. The analysis in Fig.~\ref{fel-al}e shows that the density of the domain meeting points exhibits a logarithmic-linear dependence on the degree of Al$^{3+}$ substitution in samples up to $x=0.15$. Note that the vortex density thus achieved through chemical substitution surpasses that attained by thermal quench procedures by up to an order of magnitude \cite{Griffin2012, Meier2017}.

Regarding the samples at $x>0.15$, earlier studies \cite{Sim2018} revealed a decrease of Curie temperature with increasing Al$^{3+}$ substitution, and it is possible that YMn$_{1-x}$Al$_{x}$O$_3$ at $x\geq 0.20$ is no longer ferroelectric at room temperature. On the other hand, the granular structure in the PFM image in Fig.~\ref{fel-al}e indicates that ferroelectric order may still be present, albeit with domains below the resolution limit achieved with the PFM tip. In fact, Fig.~\ref{fel-al} shows no tendencies for saturation, so that a further decrease in domain size with increasing substitution may be expected if ferroelectric order occurs in the first place.

\begin{figure}[h!] 
\centering
\includegraphics[width = \columnwidth]{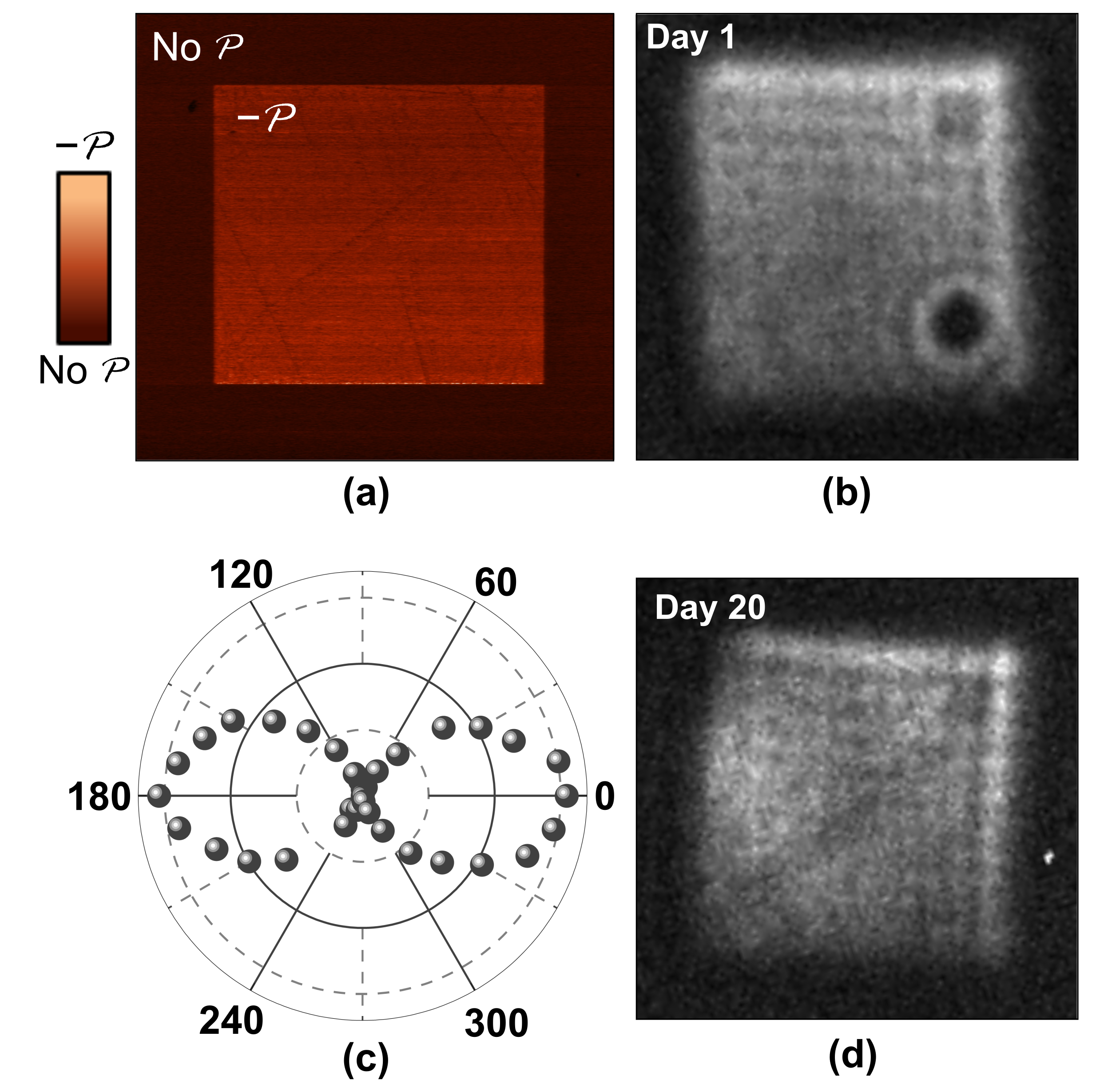}
\caption{
\label{poling} Electric-field poling of YMn$_{1-x}$Al$_{x}$O$_{3}$ at $x=0.20$. (a) Room-temperature out-of-plane PFM scans on a $c$-oriented sample after scanning the central quadratic region with the PFM tip to which 250~V were applied (E~=~3.6~MV/m). The bright part corresponds to non-zero out-of-plane polarization. The dark region is the multi-domain state with zero net polarization. (b) Spatially resolved SHG image of the region in (a). The bright part reveals the presence of an electrically net-polarized region. (The darkened circular-shaped object in the SHG image in (b) results from a dust particle). (c) SHG anisotropy measurement on the bright part in (b). The polarization of the incoming fundamental and of the detected frequency-doubled light are rotated simultaneously. The $0^{\circ}$ and $90^{\circ}$ positions correspond to the SHG components $\chi_{zzz}$ (permitted by ferroelectric order, maximum) and $\chi_{yyy}$ (forbidden by ferroelectric order, zero), respectively. (d) Like (b), but taken 20 days later. (b) and (c) use the same intensity scale, evidencing that the SHG signal and, hence, the polarization induced by the electric-field tip poling are stable. The SHG polar plot in (c) and SHG images in (b, d) are acquired at room temperature with the sample rotated by an angle of 30$^{\circ}$ between the incident fundamental light beam and the $z$ axis of the crystal.}
\end{figure}

For clarifying the state of the $x\geq 0.20$ samples, we resort to PFM tip-poling experiments. A $50\times 50$~$\mu$m$^{2}$ square is electrically poled line-by-line, applying $-250$~V between the PFM tip and the contacted rear side of the sample. The resulting out-of-plane amplitude of the PFM signal is shown in Fig.~\ref{poling}a. It reveals a distinct, uniform change in the PFM response in the poled area that is consistent with the emergence of the spontaneous net polarization characterizing a ferroelectric state. For excluding space-charge effects as origin of the enhanced PFM response, we performed supplementary SHG imaging experiments on the area probed by PFM. The spatially resolved distribution of the SHG signal in Fig.~\ref{poling}b yields uniform SHG intensity in the poled, quadratic region and zero intensity around it. In agreement with the methodical discussion in Section~\ref{methods}, this points to the presence of a uniformly polarized, inversion-symmetry-breaking ferroelectric state in the electric-field-poled region of the $x\geq 0.20$ sample. In the surrounding region, not subjected to the electric poling field, we consequently assume a multi-domain configuration with zero net polarization so that the resulting SHG signal is zero as observed. Note that according to Fig.~\ref{poling}d, the SHG image is stable for at least 20 days, which further confirms a spontaneous polarization rather than space charges as the origin of the PFM and SHG responses in the electrically poled region. Finally, a polarization analysis of the SHG signal reveals the characteristic double-lobe structure obtained from $\chi_{zzz}$ on a tilted sample (Fig.~\ref{poling}c); see the discussion of methods in Section~\ref{methods}.

\subsection{Short-range ferroelectric correlations} \label{felcor}
In order to measure the sub-PFM-resolution size of the ferroelectric domains in YMn$_{1-x}$Al$_{x}$O$_3$ at $x\geq 0.20$, we performed synchrotron diffraction measurements. This method is sensitive to the lattice-trimerizing distortion which, as discussed in Section~\ref{methods}, is coupled one-to-one to the improper ferroelectric polarization. 

\begin{figure*}[ht!] 
\centering
\includegraphics[width=133 mm]{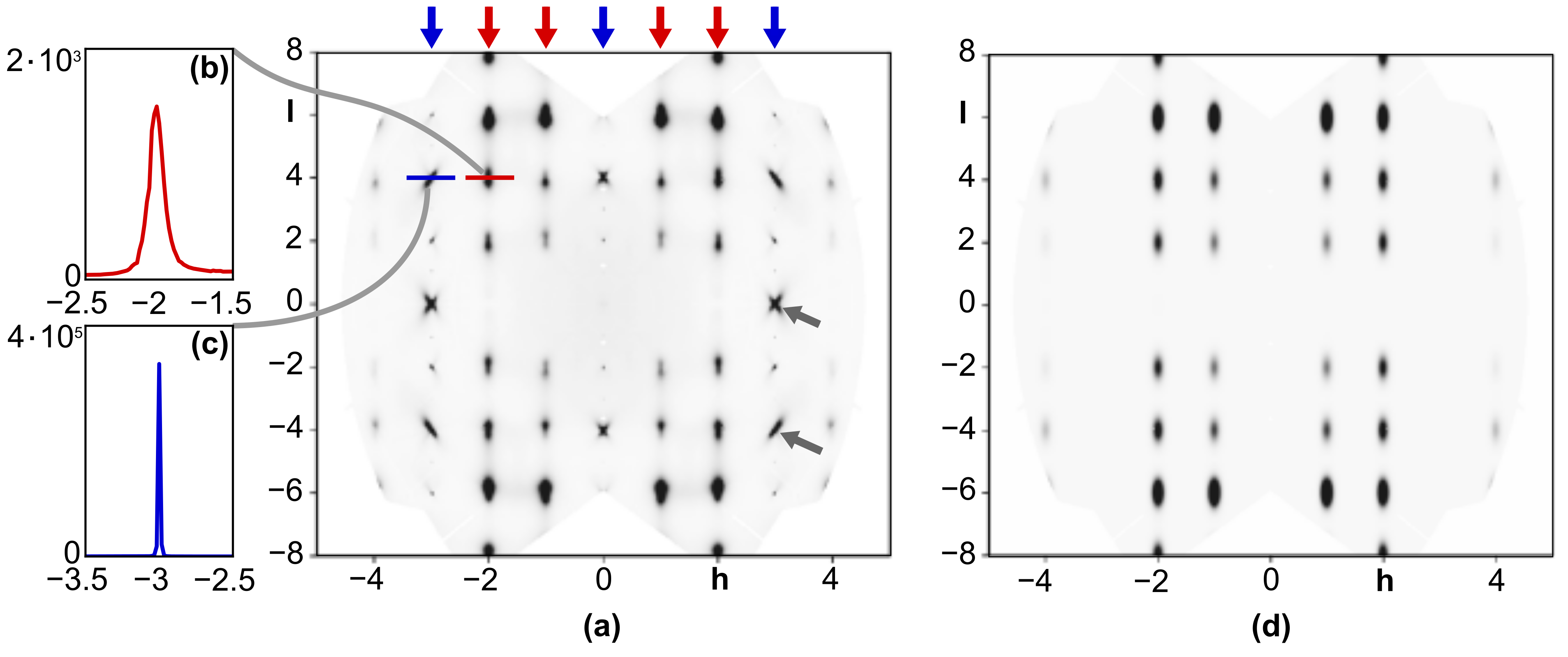}
\caption{\label{xrd} X-ray diffraction data from YMn$_{1-x}$Al$_{x}$O$_3$ at $x\geq 0.20$. (a) Experimental data in the $h0l$ plane. Broad Bragg peaks (red arrows and panel b) are associated with the structural trimerization, whereas sharp Bragg peaks (blue arrows and panel c) relate to the paraelectric parent structure. (b) and (c) are cross sections through the $\bar{2}04$ and $\bar{3}04$ Bragg peaks respectively. The broadening of the trimerized peaks reveals the ferroelectric domain correlation length of 16~\AA{} along the $c$ axis and 35~\AA{} along the $a$ axis. (d) Calculated model for the trimerization-related Bragg peaks in (a).}
\end{figure*}

Figure~\ref{xrd}a shows the diffraction $h0l$ plane of the YMn$_{1-x}$Al$_{x}$O$_3$ sample at $x=0.20$, measured at room temperature. The most prominent feature of this map is the presence of two types of Bragg reflections with very different widths. The reflections corresponding to the structural trimerization (blue arrows) are very broad. This is a typical signature of a small coherence length, which in this case means that the trimerization domains have a small size. In contrast, the Bragg peaks corresponding to the paraelectric parent phase (red arrows) are sharp, which is consistent with the excellent quality of our single crystal. Another feature of the map is the presence of butterfly-shaped diffuse scattering around certain Bragg peaks (gray arrows), which is reminiscent of Huang scattering \cite{Huang1947}. Such scattering is typically caused by lattice relaxation around point defects, such as the statistically distributed Al$^{3+}$ ions as one likely explanation.

\begin{figure}[ht!] 
\centering
\includegraphics[width = \columnwidth]{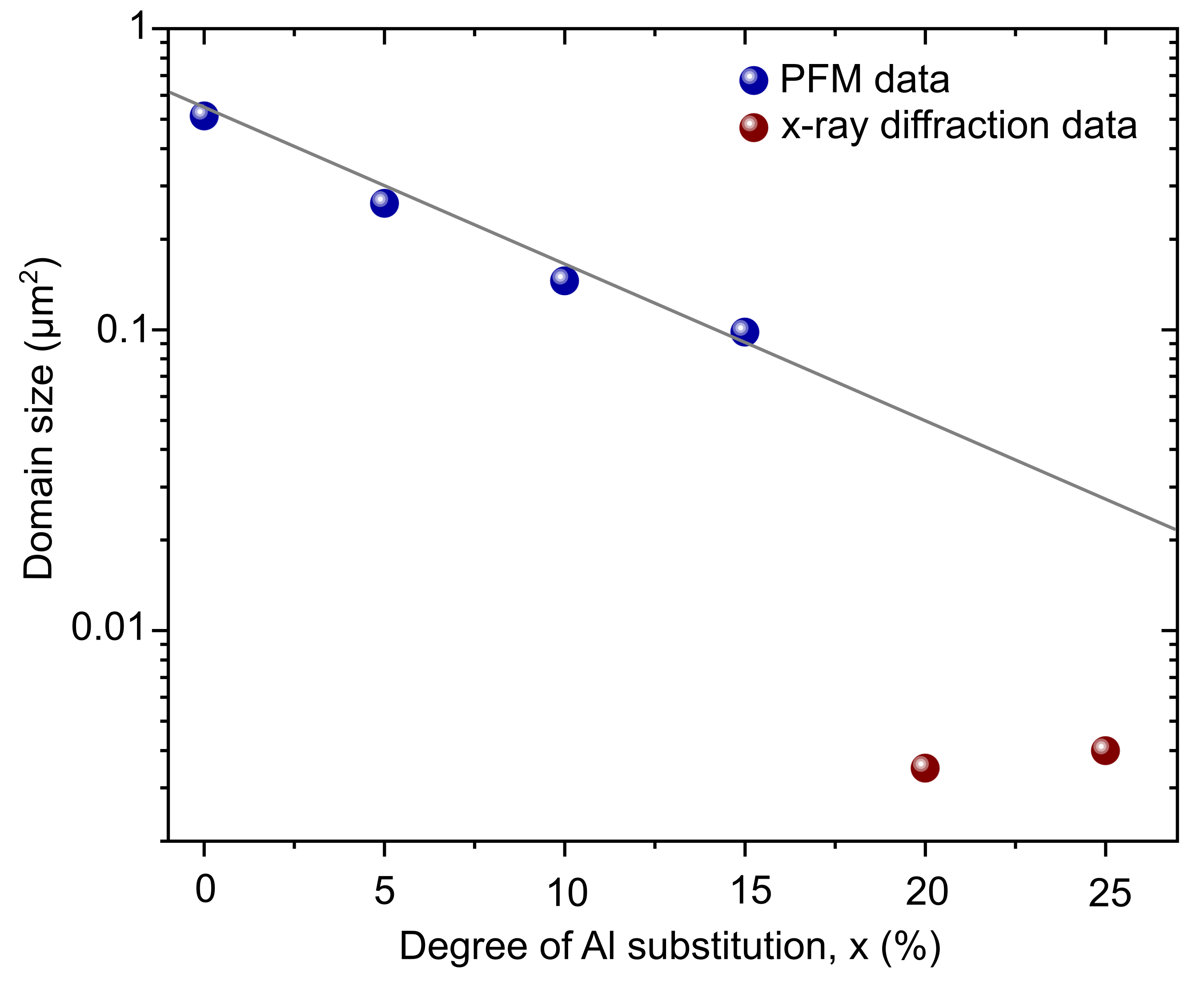}
\caption{\label{linear size} Domain size YMn$_{1-x}$Al$_{x}$O$_3$ from PFM (blue) and diffraction (red) experiments. The line is a guide to the eye.}
\end{figure}

In order to confirm that the trimerization pattern of the $x=0.20$ sample is in agreement with that of the rest of the series, we have modelled the broad Bragg peaks observed in Fig.~\ref{xrd}a. We took the experimental intensity of the corresponding peaks for the $x=0.15$ sample and convolved them with an anisotropic Gaussian function. The modelled data is presented in Fig.~\ref{xrd}d and gives an excellent agreement with the measured data. The fit of the Gaussian width leads to a correlation length of 35~\AA{} in the $ab$-plane and of 16~\AA{} along the $c$-direction. In the same way, we obtain a correlation length of 40~\AA{} in the $ab$-plane and of 20~\AA{} along the $c$-direction for the $x=0.25$ sample. To simplify comparison, we have measured the linear domain size for the rest of the series by Fourier transforming the PFM data and fitting it in the same way as diffraction data and present result in the Fig. \ref{linear size}. We therefore find that Al$^{3+}$ substitution permits us to tune the size of the ferroelectrically correlated regions from about 1~$\mu$m lateral extension at $x=0$ to about 30~\AA{} at $x=0.20$ and thus across three orders of magnitude. At $x=0.20$, the correlation-length dependence on the substitution saturates so that little further change occurs for the $x=0.25$ sample.

\subsection{Origin of the domain-size scaling}

\begin{figure*}[ht!] 
\centering
\includegraphics[scale = 1]{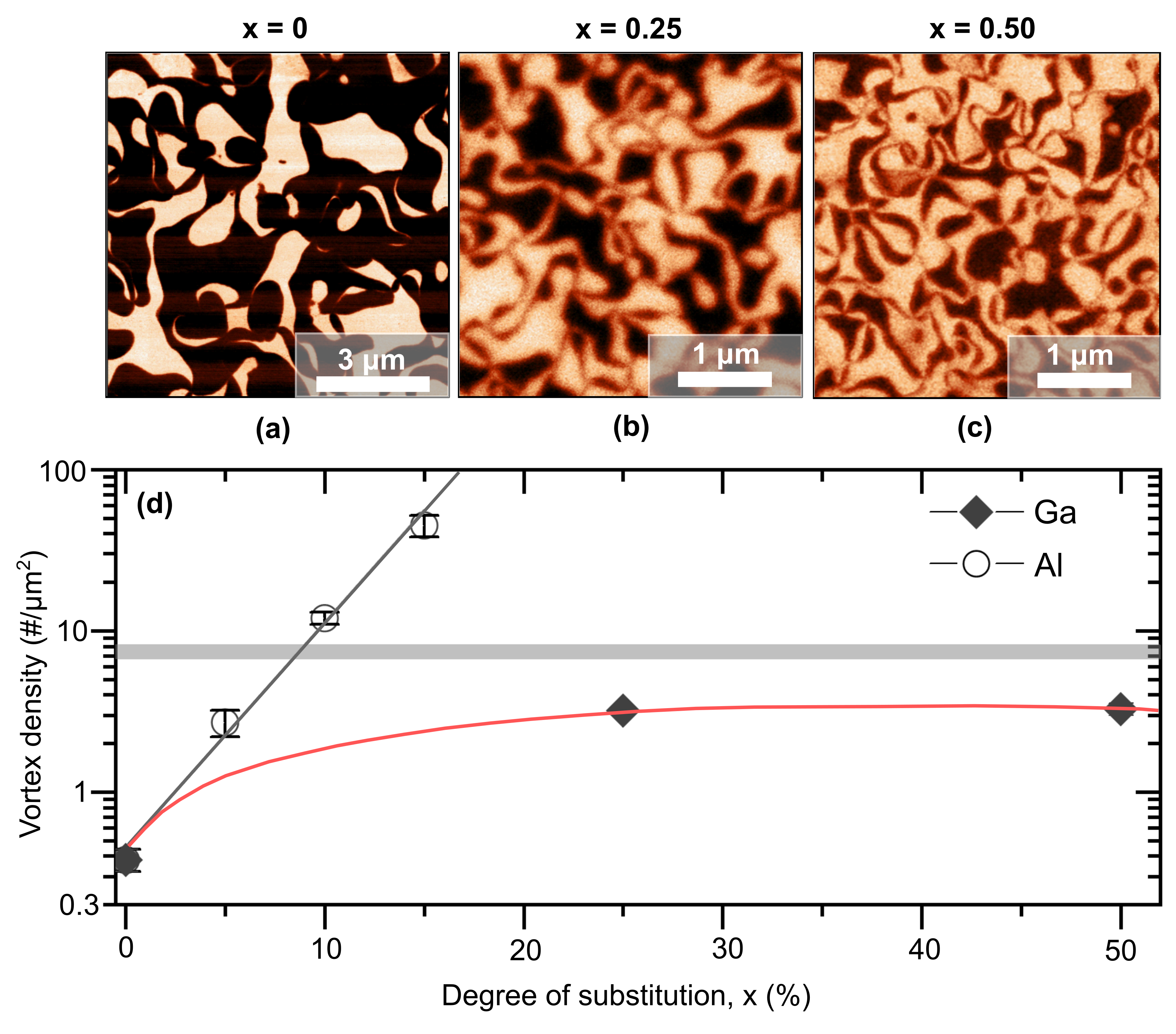}
\caption{\label{ga-fel} Ferroelectric domains and vortex densities on compounds from the YMn$_{1-x}$Ga$_{x}$O$_{3}$ series. (a -- c) Room-temperature out-of-plane PFM scans on a $c$-oriented area of (a) $10\times 10$~$\mu$m$^{2}$ and (b, c) $5\times 5$~$\mu$m$^{2}$ for different substitution $x$. (d) Density of six-fold vortex-like domain meeting points on YMn$_{1-x}$Ga$_{x}$O$_{3}$ and YMn$_{1-x}$Al$_{x}$O$_{3}$ series. The gray line represents the linear fit in Fig.~\ref{fel-al}. The red line is a guide to the eye. The gray horizontal bar indicates the largest vortex density achieved in YMnO$_3$ by thermal quenching.}
\end{figure*}

In order to benefit from the domain-size control established in Section \ref{felcor}, we need to understand its microscopic origin. One of the biggest advantages of the hexagonal manganites is their compositional versatility, and in earlier experiments, exchange or partial substitution of one of its ions revealed fundamental details about the microscopy of the system \cite{Meier2017, Sim2018, FiebigBook2023, Bieringer2002}. In the present case, we use YMn$_{1-x}$Ga$_{x}$O$_3$ as comparison system to shed light on the underlying domain-size control mechanisms. Both Al$^{3+}$ and Ga$^{3+}$ have the same valence as Mn$^{3+}$, but differ in its ionic radius, Al$^{3+}$$=$0.675~\AA, Ga$^{3+}$$=$0.76~\AA, Mn$^{3+}$$=$0.785~\AA. Figures~\ref{ga-fel}b and \ref{ga-fel}c show the ferroelectric domain pattern at room temperature in the $ab$ plane of two YMn$_{1-x}$Ga$_{x}$O$_3$ samples with increasing Ga$^{3+}$ substitution. In stark contrast to the Al$^{3+}$-substituted samples, the Ga$^{3+}$-based samples in Figures~\ref{ga-fel}b and \ref{ga-fel}c exhibit a significantly reduced dependence of the ferroelectric domain size on the degree of substitution, despite the much larger degree of substitution that is reached in comparison to the Al$^{3+}$-based samples. The analysis in Figure~\ref{ga-fel}d shows that the density of domain meeting points increases from $x=0$ to $x=0.25$ and remains constant from $x=0.25$ to $x=0.50$. In particular, it does not even reach the value attained by thermal-quench experiments \cite{Griffin2012, Meier2017}.

\begin{figure*}[ht!] 
\centering
\includegraphics[scale = 1]{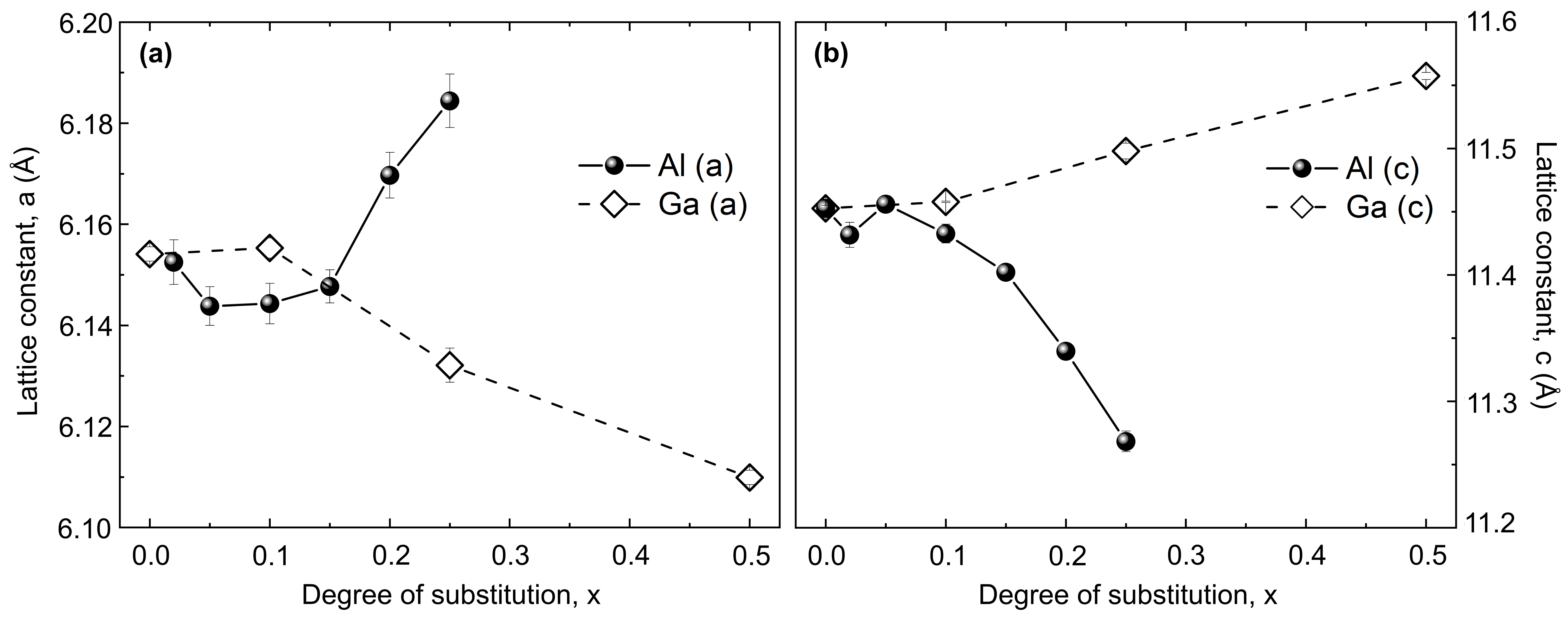}
\caption{\label{distance} (a) $a$ and (b) $c$ lattice constants in the YMn$_{1-x}$Al$_{x}$O$_{3}$ and YMn$_{1-x}$Ga$_{x}$O$_{3}$ series. The data sets used in these plots are taken from Ref.\,\cite{Sim2018}.}
\end{figure*}

In the literature we find several indications that the size of the ions and the resulting lattice distortions have an influence on central physical aspects of the hexagonal $R$MnO$_3$ system, such as phase diagrams and transition temperatures \cite{Fiebig2003}. In Fig.~\ref{distance} we therefore compare the lattice constants of the YMn$_{1-x}$Al$_{x}$O$_3$ and YMn$_{1-x}$Ga$_{x}$O$_3$ systems as derived from our x-ray diffraction experiments. We find that Al$^{3+}$ and Ga$^{3+}$ substitutions have opposite effects on the lattice constants. Whereas we have a decrease in $c$ and an increase in $a$ with the former, we get an increase in $c$ and a decrease in $a$ with the latter.

A possible interpretation of Figs.~\ref{ga-fel} and \ref{distance} might include two mechanisms affecting the size of the ferroelectric domains. \textit{First}, according to preliminary calculations by density functional theory \cite{ASimonov} the replacement of Mn$^{3+}$ is likely to lower the anisotropy in the direction of the bipyramidal MnO$_5$ tilt. The barrier away from their preferred radial tilt (towards the Y$^{3+}$ ions) becomes lower and with it the energy penalty of the trimerization-polarization domain walls in which the tilt direction reorients. Domain walls are thus formed more easily which can contribute to the decreasing size of the ferroelectric domains with the ongoing replacement of Mn$^{3+}$ by Al$^{3+}$ or Ga$^{3+}$. 

\begin{figure*}[ht!] 
\centering
\includegraphics[scale = 1.2]{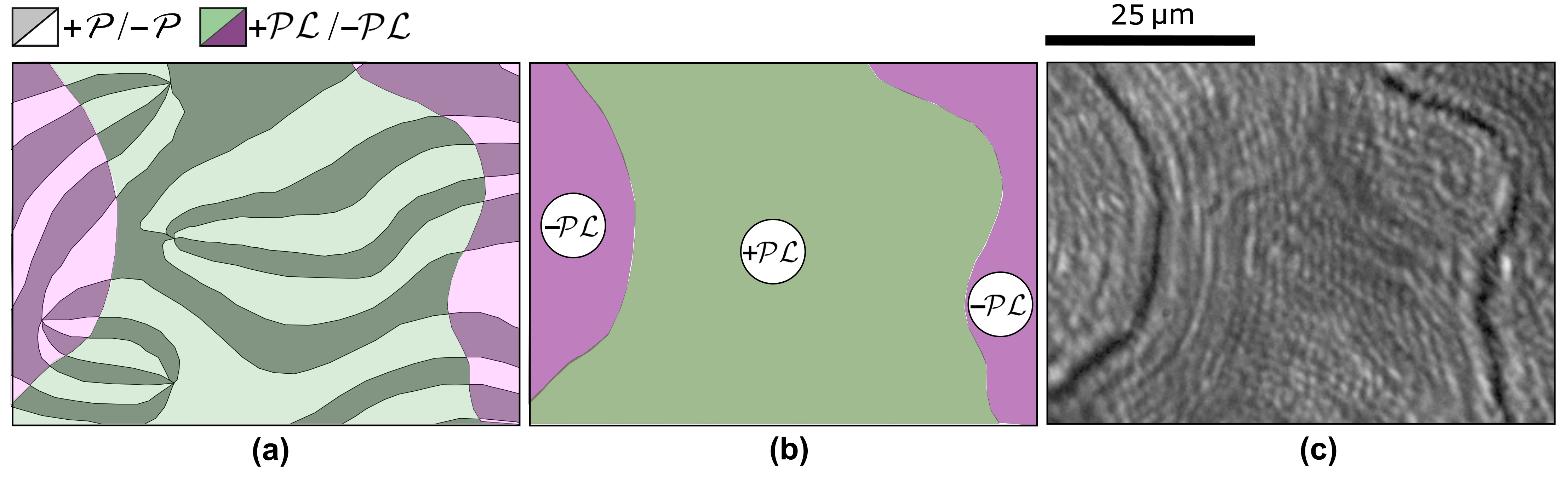}
\caption{\label{me-domains} Sketch of the coupling between ferroelectric and antiferromagnetic domains and domain walls. (a) Schematic distribution of the so-called correlated and uncorrelated antiferromagnetic domain walls (see text). The sign of $\mathcal{L}$ flips at every domain wall. (b) Sketch and (c) SHG image probing the distribution in (a) with the $\mathcal{PL}$-related contribution to SHG. The correlated antiferromagnetic walls are invisible because at these, both $\mathcal{P}$ and $\mathcal{L}$ change their sign so that $\mathcal{PL}$ and the related SHG light field remain unchanged. Only the uncorrelated antiferromagnetic walls are visible because of the sign change of $\mathcal{L}$ only and, hence, of the product $\mathcal{PL}$. As explained in the text, this leads to destructive SHG interference at the uncorrelated walls which become visible as dark grooves.}
\end{figure*}

\textit{Second}, with the change in $c$ and $a$ by Al$^{3+}$ substitution, the system approaches the lattice constants of the paraelectric phase \cite{Sim2018}. This reduces the energy cost of forming ferroelectric domain walls \cite{Selbach2019} so these are formed more easily and domains get smaller, until at $x\geq 0.20$ the entire order becomes short-range as observed. In contrast, the change in $c$ and $a$ by Ga$^{3+}$ stabilizes the ferroelectric state and can thus counteract the first mechanism above with the result of a significantly reduced dependence of the domain size on the degree of Mn$^{3+}$ replacement.
Finally, with the small value of the correlation length found for the YMn$_{1-x}$Al$_{x}$O$_3$ samples at $x\geq 0.20$, it may be interesting to discuss these in the light of relaxors \cite{Blinc2011, Woo2002}. It is proposed that the relaxor behavior is caused by the existence of polar nano-regions surrounded by the non-polar matrix. We may suspect that our crystals are likewise composed of ferroelectric domains surrounded by a paraelectric matrix. This model can be justified by the fact that the overall intensity of trimerization peaks in the $x=0.20$ and $x=0.25$ samples is weaker than in the $x=0.15$ sample by factors of 0.6 and 0.4, respectively. In the two-phase relaxor model \cite{Cowley2011} this would correspond to 60\% or 40\% ferroelectric material within a paraelectric matrix for $x=0.20$ and $x=0.25$, respectively. Here the paraelectric matrix may be promoted by an inhomogeneous distribution of the Al$^{3+}$ substitutes with a lower and higher local concentration corresponding to ferroelectric and paraelectric nanoregions, respectively. An alternative explanation might be that the paraelectric regions are associated with the domain walls which are present in the $x\geq 0.20$ samples in an extreme concentration. For ultimately relating the highly substituted YMn$_{1-x}$Al$_{x}$O$_3$ samples to relaxors, one would have to measure their frequency-dependent dielectric response. Unfortunately this is not possible at present because of the inferior sample quality so that the relaxor nature has to remain a matter of speculation for now.

\subsection{Magnetoelectric correlations}

In the final step, we scrutinize how the chemical substitution on the Mn$^{3+}$ site affects the coupling between ferroelectric and antiferromagnetic domains in the YMn$_{1-x}$Al$_{x}$O$_{3}$ system. The magnetoelectric coupling relates to the different types of domain walls that occur in the $R$MnO$_3$ family \cite{Giraldo2021}, so that an overview of these is given in Fig.~\ref{me-domains}a. We see that the majority of domain walls are spatially coinciding ferroelectric-antiferromagnetic walls. The associated so-called \textit{correlated antiferromagnetic domain walls} reproduce the vortex-like domain pattern of the ferroelectric state. This network is complemented by a smaller fraction of so-called \textit{uncorrelated antiferromagnetic domain walls}, which are not coupled to any ferroelectric walls. They form a pattern of curvy, isotropic domains without any vortex-like features. 

\begin{figure*}[ht!] 
\centering
\includegraphics[scale = 1]{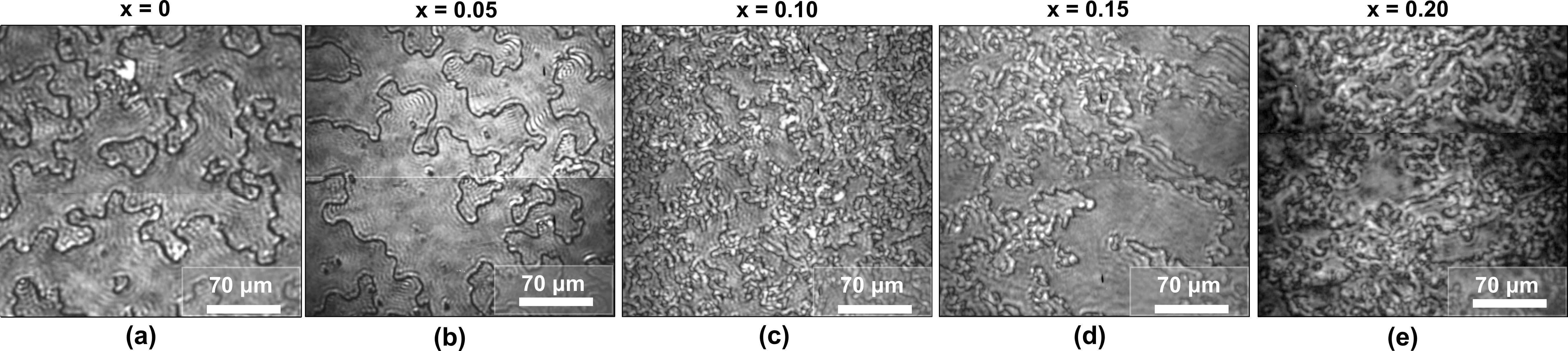}
\caption{\label{afm-domains} Domains and uncorrelated antiferromagnetic domain walls (dark grooves, see Fig.~\ref{me-domains}) at 20~K on $c$-oriented samples from the YMn$_{1-x}$Al$_{x}$O$_{3}$ series. The observation of the $\mathcal{PL}$ domains throughout the substitution series from $x$=0 to $x$=0.20 shows that the magnetoelectric coupling between ferroelectric and antiferromagnetic domains is retained (see text).}
\end{figure*}

Note that the SHG contribution from $\mathcal{PL}$ mentioned in Section~\ref{methods} probes the distribution of the uncorrelated antiferromagnetic walls. At these walls, $\mathcal{PL}$ experiences a sign change, leading to a sign change in the emitted SHG light field \cite{Giraldo2021, Fiebig2002}. This corresponds to a $180^{\circ}$ phase difference between SHG contributions and leads to destructive interference in the vicinity of the uncorrelated antiferromagnetic walls. Because of this, these walls appear as dark grooves in spatially resolved SHG images. In contrast, at the magnetoelectrically coupled correlated antiferromagnetic domain walls, both $\mathcal{P}$ and $\mathcal{L}$ change simultaneously so that no change in the $\mathcal{PL}$-related SHG signal is expected. This leads to the unusual situation that the \textit{absence of the vortex-like domain structure in the SHG images representing the $\mathcal{PL}$ distribution indicates that the magnetoelectric coupling of the $\mathcal{P}$ and $\mathcal{L}$ domains and domain walls is still intact}. This concept is also sketched in Fig.~\ref{me-domains}b.

After these introductory remarks, we can interpret the spatially resolved $\mathcal{PL}$-related distribution of SHG that is depicted for a series of YMn$_{1-x}$Al$_{x}$O$_{3}$ samples in Fig.~\ref{afm-domains}. Only the samples from $x=0$ to $x=0.20$ are shown because at the acquisition temperature of 20~K, the sample at $x=0.25$ is still in the paramagnetic phase. For all compositions $x$ we observe an approximately flat distribution of SHG intensity that is superimposed by a distribution of curvy dark grooves. As discussed before, these grooves indicate the position of the uncorrelated antiferromagnetic domain walls of the samples. Since no indication of a vortex-like distribution of domains is observed, we conclude that, as mentioned above, $\mathcal{P}$ and $\mathcal{L}$ remain correlated as sketched in Fig.\ref{me-domains}a, and thus the magnetoelectric coupling of the $\mathcal{P}$ and $\mathcal{L}$ domains and domain walls remains intact in all these samples. Considering the high degree of substitution and the transition to short-range trimerized-polar order with a correlation length of barely a few unit cells in the $x=0.20$ sample, it is quite surprising that the characteristic magnetoelectric coupling mechanism of the $R$MnO$_3$ series is not affected. In stark contrast to the density of the ferroelectric domain walls, the density of the uncorrelated antiferromagnetic domain walls does not change drastically across the substitution series.


\section{CONCLUSIONS}

In summary, substitution of Mn$^{3+}$ by Al$^{3+}$ in YMnO$_{3}$ as model representative of the hexagonal multiferroic $R$MnO$_3$ family is a way for tuning the lateral size of the ferroelectric domains across three orders of magnitude. In a combination of PFM, optical SHG, and synchroton x-ray diffraction experiments we find that the expansion of the uniformly polarized regions in the YMn$_{1-x}$Al$_{x}$O$_{3}$ system ranges between a few micrometers at $x=0$ and a few unit cells at $x=0.25$. At $x=0.20$ the distortive lattice trimerization inducing the improper ferroelectric order becomes short-range and reaches its lower limit. Anisotropic strain accompanying the Mn$^{3+}$ replacement may explain the domain-size variation. Even though the distortive-polar lattice undergoes these drastic changes, the magnetoelectric coupling of the ferroelectric and antiferromagnetic domains and domain walls remains untarnished. The distance between the Y$^{3+}$ ions along the hexagonal axis is identified as origin of the size variation of the ferroelectric domains. In combination with established thermal annealing procedures \cite{Griffin2012, Meier2017}, the tuning range for the domain size may be further expandable. 

We thus find that chemical substitution is a very effective means of domain-size engineering in one of the most investigated multiferroics. Since the ferroelectric properties exhibit little sensitivity to the choice of $R^{3+}$ ion, we assume that our observations on YMnO$_3$ are representative for the $R$MnO$_3$ family at large.


\section*{Acknowledgments} We acknowledge the X-Ray Platform at the Department of Materials at ETH Zurich for the provision of their single-crystal diffractometer for preliminary x-ray measurements. M.G. and T.L. thank Quintin Meier from Institut N\'{e}el CNRS for fruitful discussions about the correlation between the structural parameters and average domain size in the family of hexagonal manganites. M.G. and L.F. thank Mads C. Weber for advice on handling and preparing brittle samples. The authors also thank Marta Rossell from the Electron Microscopy Center at Empa for preliminary transmission electron microscopy measurements. A.S. thank the European Synchrotron Radiation Facility (ESRF) for provision of their equipment and we would like to thank Artem Korshunov for assistance and support in using beamline ID28. M.G., A.S., L.F., T.L., and M.F. received funding by the Swiss National Science Foundation through the grants No.\ 200021--215423, PCEFP2--203658 and 200021--178825 and the EU European Research Council (Advanced Grant 694955--INSEETO). M.T. acknowledges the financial support by the Swiss National Science Foundation under project No.\ 200021--188414. The work at SNU was supported by the Leading Researcher Program of the National Research Foundation of Korea (Grant No. 2020R1A3B2079375). \\

\section*{Author Contributions} M.G. designed, conducted the SHG experiments, and analyzed the SHG data supervised by T.L. and M.F. The synchrotron diffraction experiments were designed, conducted, and analyzed by A.S. The single crystals were grown by H.S. under the supervision of J.-G.P. Furthermore, L.F. and M.G. pre-oriented the crystals using x-ray single-crystal diffraction and prepared the samples. M.L. performed the room-temperature PFM measurements on the YMn$_{1-x}$Al$_{x}$O$_{3}$ samples. A.S.L. performed the PFM poling experiment on YMn$_{0.80}$Al$_{0.20}$O$_{3}$. E.G. performed the room temperature PFM measurements on the YMn$_{1-x}$Ga$_{x}$O$_{3}$ samples. All experimental results were discussed and interpreted by M.G., T.L., M.T., A.S., and M.F. The manuscript was written by M.G., T.L., A.S., and M.F. with input and discussion by all authors.


\newpage


\newpage

\setcounter{figure}{0}
\renewcommand{\thefigure}{S\arabic{figure}}

\end{document}